\documentstyle [12pt]{article}    
\def\gsimeq
{\hbox{\raise0.5ex\hbox{$>\lower1.06ex\hbox{$\kern-1.07em{\sim}$}$}}}
\def\lsimeq
{\hbox{\raise0.5ex\hbox{$<\lower1.06ex\hbox{$\kern-1.07em{\sim}$}$}}}
\def\pn{\par\noindent}

\def\ms{\medskip\pn}
\def\bs{\bigskip\pn}

\begin{document}

\renewcommand{\thefootnote}{\fnsymbol{footnote}}

\title{Warm absorber, reflection and Fe K line in
the X-ray spectrum of IC 4329A}

\author{M. Cappi$^{1}$, T. Mihara$^{1}$, M. Matsuoka$^{1}$,
K. Hayashida$^2$, \\
K.A. Weaver$^{3,4}$ and C. Otani$^{1}$}
\date{ }

\maketitle
\pn
   $^1$ The Institute of Physical and Chemical Research (RIKEN),
2-1, Hirosawa, Wako, Saitama 351-01, Japan
\pn
   $^2$ Department of Physics, Faculty of Science, University of
Osaka, 1-1 Machikaneyama-chou, Toyonaka, Osaka 560
\pn
   $^3$ Department of Astronomy and Astrophysics, 525 Davey Lab.,
Pennsylvania State University, University Park, PA 16802
\pn
   $^4$ Presently at The Johns Hopkins University, Department of Physics
and Astronomy, 3400 N. Charles Street, Baltimore, MD 21218
\pn
\pn
\bs
\pn
\pn
\pn
\pn
\pn
\pn
\pn
\pn

\vfill\eject

\begin{center}
{\large \bf ABSTRACT}
\end{center}
Results from the X-ray spectral analysis of the ASCA PV phase observation
of the Seyfert 1 galaxy IC 4329A are presented. We find that the 0.4 - 10 keV
spectrum of IC 4329A is best described by the sum of a steep
($\Gamma \sim 1.98$) power-law spectrum
passing through a warm absorber plus a strong reflection component and
associated Fe K line, confirming recent results (Madejski et al. 1995,
Mushotsky et al. 1995). Further cold absorption in excess of the Galactic
value and covering the entire source is also required by the data, consistent
with the edge-on galactic disk and previous X-ray measurements.
The effect of the warm absorber at soft X-ray energies
is best parameterized by two absorption edges, one consistent with OVI, OVII
or NVII, the other consistent with OVIII. A description of the soft
excess in terms of blackbody emission, as observed in some other Seyfert 1
galaxies, is ruled out by the data.
A large amount of reflection is detected in both the GIS and SIS detectors,
at similar intensities.
We find a strong correlation between the amount of reflection
and the photon index, but argue that the best solution with the present
data is that given by the best statistical fit. The model dependence
of the Fe K line parameters is also discussed.
Our best fit gives a slightly broad ($\sigma \simeq 0.11 \pm 0.08$ keV) and
redshifted (E $\simeq 6.20 \pm 0.07$ keV) Fe K line, with equivalent
width $\simeq$ 89 $\pm$ 33 eV. The presence of a weak Fe K line with a strong
reflection can be reconciled if one assumes iron underabundances or ionized
reflection.
We also have modeled the line with a theoretical line profile produced by an
accretion disk. This yields results in better agreement with the constraints
obtained from the reflection component.

\bigskip
{\bf Key-words}: galaxies: individual (IC 4329A) - galaxies: Seyfert -
X-rays: galaxies

\vfill\eject

\section{Introduction}

X-ray observations of Seyfert 1 galaxies strongly support the
hypothesis that optically thick (column density $\gsimeq\ 10^{23}$ cm$^{-2}$)
matter
surrounds the innermost regions of Active Galactic Nuclei (AGN) (Nandra \&
Pounds 1994).
The detection of a bump in the continuum emission above $\sim$
7 keV and an iron fluorescence line at $\sim$ 6.4 keV in several
Seyfert galaxies are compelling evidence
that the intrinsic continuum is reprocessed by cold matter (Guilbert \& Rees
1988, Lightman \& White 1988).
Recent ASCA observations also confirm the existence of ionized matter (the
so-called warm absorber) along the line of sight of some Seyfert 1 galaxies
(e.g., MCG-6-30-15: Fabian et al. 1994; NGC3227: Ptak et al. 1994; NGC3783:
George, Turner \& Netzer 1995).
In these sources, the warm absorber alone
can explain both the absorption features detected around 1 keV and
the soft-excess emission. However, this seems not to be the rule for all
Seyfert 1 galaxies (Matsuoka 1994).
Some of them, like NGC7469, do not show the presence of ionized
absorption but require extra
soft emission best characterized by a low temperature blackbody (Guainazzi
et al. 1994). Others, like NGC4051 and Mkn766, require both the warm absorber
and blackbody component (Mihara et al. 1994, Leighly et al. 1995,
Guainazzi et al., in preparation).

In light of these general considerations, the ASCA observation of the
bright Seyfert 1 galaxy IC 4329A (z=0.016, Wilson \& Penston 1979)
provides new insights on these issues. From the $Ginga$ observation in the 1.7
- 30 keV
energy band, IC 4329A is known to exhibit a strong reflection component which
flattens the 1.7 - 30 keV photon
index from an intrinsic value of $1.97 \pm 0.04$
to the apparent value of $1.71 \pm 0.01$
(Miyoshi et al. 1988, Piro, Yamauchi \& Matsuoka 1990, hereafter PYM90).
The reflection component is accompanied
by an iron fluoresence line at E $\sim$ 6.2 keV, with an equivalent width (EW)
of $\simeq 100 \pm 20$ eV. Evidence of day to day ($\sim$ 50\%) variations in
the amount of reflection relative to the direct continuum have also been
reported (Fiore et al. 1992).
In the ROSAT soft energy band (0.1-2.4 keV), the spectrum of IC 4329A
is best described by an absorbed power law with a photon index $\Gamma
\simeq 1.83 \pm 0.4$ plus an absorption edge around 0.7 keV having an optical
depth of $\simeq 0.6
\pm 0.2$ consistent with absorption by OVI and OVII (Madejski et al. 1995).
However, an alternative explanation of the soft excess in terms of a
black-body or a steep power law is also allowed by the same data.
All fits derive a column density of a few
10$^{21}$ cm$^{-2}$. This column density is largely in excess of the
Galactic value $N_{\rm Hgal} \simeq 4.55\times 10^{20}$ cm$^{-2}$ (Elvis,
Lockman \& Wilkes 1989) and indicate absorption intrinsic to IC 4329A.
This result is consistent with the fact that its host galaxy is seen edge-on
(Petre et al. 1984).

In the following, we present the results from the X-ray observation of IC 4329A
with ASCA (Tanaka, Inoue \& Holt 1994). The data between 2 - 10 keV have been
previously discussed in a letter by Mushotsky et al. (1995).
We now report a more detailed analysis of the overall (0.4-10 keV) data.
We show very strong evidence
that the X-ray spectrum of IC 4329A is absorbed by highly ionized gas (warm
absorber),
characterized by the absorption edges of highly ionized oxygen ions (section
3.1).
We also confirm the presence of a spectral hardening above $\sim$ 6 - 7 keV, as
well
as the iron K-shell emission line, which are characteristics
of a reflection spectrum (section 3.2).
Complex warm absorber models and accretion-disk emission
line models are then discussed (section 4).
We conclude that there is evidence of a warm absorber, a strong reflection
component and a marginally broad iron fluoresence line in the X-ray spectrum
of IC 4329A.

\section{The X-ray data}

The X-ray observation of IC 4329A reported here was carried out with ASCA on
August 15th 1993 during the performance verification (PV) phase.
Solid state imaging spectrometer (SIS) data
were collected in 4-CCD 'Bright' and 'Faint' modes. Faint data were
converted into Bright data. No echo corrections nor Dark Frame Error (DFE)
corrections were applied (see Otani \& Dotani 1994 for details on these
effects).
Standard selection criteria were used, the most relevant being
elevation and bright Earth angles greater than 5$^{\circ}$ and
25$^{\circ}$, respectively, and
a magnetic cut-off rigidity greater than 8 GeV/c.  These were applied
to both GIS and SIS data.
The resulting effective exposure times are $\sim$ 29000 s for the SIS and
$\sim$ 32500 s for the GIS. About 55000 and
90000 source plus background counts were
collected from each SIS and GIS, respectively.
The background count rates were about 1\% of the SIS average source count
rate ($\sim$ 1.7 counts/s) and about 5\% of the GIS average source
count rate ($\sim$ 2.6 counts/s).

Only chip n.1 of SIS0 and chip n.3 of SIS1 were used in the SIS analysis.  This
results
in the loss of $\sim$ 30 \% of the total SIS source counts
to the neighboring chips due to the broad wings of
the mirror point response function.
Source counts were extracted from circular
regions of radius $\sim$ 12 arcmin and $\sim$ 6 arcmin for the
gas imaging spectrometer (GIS) and SIS, respectively.
Because the source is bright and fills the chip,
SIS background was extracted
from the SIS blank sky files from a same area as that used for the source.
GIS background counts were accumulated from annular regions
around the source, with an internal radius of $\sim$ 14 arcmin
and an external radius of
$\sim$ 17 arcmin. Results with blank sky GIS backgrounds are consistent with
the
reported one to within $\sim$ 5\%.
It has been reported from ROSAT observations (Madejski et al. 1995) that
two other X-ray sources (the Elliptical Galaxy IC4329 and an unidentified
source) lie within $\sim$ 13 arcmin of IC 4329A.
In the following analysis, we will
assume that their contribution is negligible since they become important only
below $\sim$ 0.5 keV, as pointed out by Madejski et al. (1995).
Neither source is spatially resolved with the SIS nor with the GIS, even when
we
considered only the soft energies (E $\lsimeq$ 0.5 keV) or only the
hard energies (E $\gsimeq$ 5 keV).

Although the source flux decreased by $\sim$ 15\% in
a timescale of $\sim$ 1 day in all
detectors, a hardness ratio analysis reveals no
significant spectral variation during the observation. Therefore, all
photons have been accumulated together for the spectral
analysis.

\section{Spectral analysis}

GIS2/3 and SIS0/1 pulse height spectra were binned in order to have a number
of counts/bin  $\gsimeq$ 200 and $\gsimeq$ 100 in the energy ranges
(0.7-10 keV) and (0.4-10 keV), respectively.
Data preparation and spectral analysis have been performed using
version 1.0h of the XSELECT package and version 8.5 of the XSPEC program
(Arnaud et al. 1991). We made use of the response matrices gisv3$\_$1.rmf
(released in March 1994) for the GIS detectors and rsp1.1alpha (released
in June 1994) with Dark Frame Error (DFE)
shifted by -2 ADU ($\sim$ 7 eV) and -5 ADU ($\sim$ 18 eV) for the
SIS0 and SIS1 detectors, respectively.

For the GIS data, we first modelled the overall (0.7 - 10 keV) spectrum with a
single
absorbed power law plus
iron emission line. The column density, $N_{\rm H}$, was free to vary.
This model gives an acceptable ($\chi^{2}_{red}$/d.o.f.
$\simeq$ 1.33/564) description of the spectrum
with $\Gamma \simeq 1.79 \pm 0.02$, a column density $N_{\rm H} \simeq
(3.8 \pm 0.2)\times 10^{21}$cm$^{-2}$ and a very broad ($\sigma \simeq 0.53
\pm 0.24$ keV) iron emission line at
E $\simeq$ 6.19 $\pm 0.15$ keV with EW $\simeq$ 233 $\pm\ 80$ eV
(see Table 1A).
With these values, the unabsorbed flux in the 2 - 10 keV energy
band is $\simeq 1.01 \times 10^{-10}$ erg cm$^{-2}$ s$^{-1}$,
which corresponds to a luminosity L$_{2-10\rm keV} \sim$ 1.07
$\times$ 10$^{44}$ erg s$^{-1}$.
Although the continuum shape is approximately consistent with previous
measurements with the same model (PYM90, Madejski et al. 1995), the iron line
appears much broader and a factor of $\sim$ 2 stronger
than previous values (PYM90, Nandra \& Pounds 1994).
However, this difference can be easily understood if we consider that, as
shown later on, the SIS data strongly requires the addition of a reflection
component to the continuum emission (section 3.2). Because of its better
resolution, the SIS can indeed better separate the Fe K line from the
reflection component. When the reflection is included in the model, the
Fe K line parameters obtained from the GIS data are considerably reduced
(Table 1B) and become consistent with $Ginga$ and SIS results.
A detailed description
of the reflection parameters given in Table 1B, as well as the full
justification for the use of this model, is presented in section 3.2, in the
light of the results obtained from the SIS spectrum.

For the SIS data, a single absorbed power law
model with $N_{\rm H} \sim$ 4 $\times$ 10$^{21}$ cm$^{-2}$ and
$\Gamma \sim$ 1.75 gives an unacceptable fit, with $\chi^{2}_{red}$/d.o.f.
$\simeq$ 2.21/551. The residuals to this fit between 0.4 - 10 keV are
shown in Figure 1 and are indicative of the complexity of the X-ray
spectrum of IC 4329A. At soft energies, the
residuals show the presence of excess emission at E $\lsimeq$ 0.8 keV
and absorption features around 1 keV.
These indicate either that the underlying soft continuum emission is more
complex than a single power law or that the absorption is more complex
than a uniform, cold absorption.
At higher energies, we find an Fe K emission line at E $\sim$ 6.2 keV but
also systematic positive residuals starting from E $\gsimeq$ 5 keV. The
presence of this ``hard tail'' will appear more evident (section 3.1.2.)
after fitting adequately the soft energy part of the spectrum.

Because of the complexity of the SIS spectrum, we split the following
analysis in two parts: (i) fit of the spectrum between 0.4 - 5 keV in order to
focus only into the characteristics of the soft energy band, (ii) fit of
the overall (0.4-10 keV) spectrum.

\subsection{Soft energy band}

\subsubsection{Thermal component}

One possibility to explain the measured soft excess would be to assume a
two component continuum emission instead of a single power law. A steep,
ultra soft excess may be associated with emission from the inner
parts of an accretion disk
(Wilkes \& Elvis 1987, Turner and Pounds 1989).
As a first approximation, this soft emission can be parameterized
by a low temperature (T $\lsimeq$ 10$^6$ K, kT $\lsimeq$ 100 eV) blackbody.
A model consisting of a blackbody plus absorbed power law was fitted to the
0.4 - 5 keV SIS spectrum (SIS0 and SIS1 are fitted simultaneously).
The resulting blackbody temperature of $\sim 9.3 \times 10^5$ K (kT $\sim$ 80
eV),
photon index of
$\sim$ 1.96 and column density of $\sim\ 5.8 \times 10^{21}$ cm$^{-2}$ are
consistent with previous ROSAT results (Madejski et al. 1995), but
the residuals shown in Figure 2$a$ present strong and systematic
deviations around E $\simeq$ 0.6 keV and absorption features around 0.9 keV.
These clearly indicate that the blackbody model cannot, alone, explain
the measured soft excess.
The further addition of an absorption edge at E $\sim$ 0.76 keV with
$\tau \sim$ 0.46 improves the statistics of the fit to
$\chi^{2}_{red}$/d.o.f. $\simeq$ 1.22/481. In that case,
the normalization of the blackbody is reduced by a factor of $\sim$ 5 and
some systematic structures are still present in the residuals around 0.9 keV.
These imply that also this possibility is unlikely.

Alternatively, we modelled the soft excess with a Raymond-Smith plasma model
(Raymond \& Smith 1986) absorbed by the Galactic column only.
This model has been proposed in the past as a
possible explanation for the soft excess observed in some Seyfert 1
galaxies (Turner et al. 1991).
The best result is obtained allowing the abundances to be free. This gives a
plasma
temperature of $\sim 2.1 \times 10^6$ K (kT $\sim$ 183 eV) and improves the fit
to
$\chi^{2}_{red}$/d.o.f. $\simeq$ 1.28/479.
However, the best-fit abundances are $\sim$ 1/20 the solar value.
Such a required low abundance is obvious
from the absence of any blend of emission lines around 0.8-0.9 keV.
Moreover, some systematic features
are present in the residuals (Figure 2$b$). Thus we also exclude this
alternative as a possible description of the soft X-ray spectrum of IC 4329A.

\subsubsection{Absorption edges}

Another possibility to explain the soft excess is
an ionized or "warm" absorber.
The ionized absorber model assumes that light elements (O, Ne, Mg, Si, Fe) are
photoionized and therefore predicts extra emission due to the
reduction of the opacity at soft energies.  This model
also predicts extra absorption
around 1 keV, mainly due to highly ionized oxygen (Netzer 1993).
As clear from the residuals shown in Figure 1, both
these features are detected in the spectrum of IC 4329A. These strongly
support,
a priori, the ionized absorber model.

As a first approximation of a warm absorber, we added a single
absorption edge to
the absorbed power law model. Given the good statistics of the present
data ($\sim 50000$ counts/SIS) and the ASCA spectral resolution
(FWHM $\sim$ 80 eV at 1 keV), a single absorption edge is a rather crude
approximation of a warm absorber, but it allows a direct comparison with
previous measurements.
The inclusion of an absorption edge in the absorbed power law model
improves the fit to $\chi^{2}_{red}$/d.o.f. $\simeq$ 1.30/479 and
gives E$_{edge} \simeq 0.76 \pm 0.01$ keV and $\tau_{edge} \simeq 0.84
\pm 0.04$.
These values are consistent with the ROSAT results (Madejski et al. 1995).
However, inspection of the residuals in Figure 2$c$ clearly suggests
further complexity. Indeed, the addition of a second
absorption edge improves the statistics to $\chi^{2}_{red}$/d.o.f. $\simeq$
1.15/477, which is significant at more than 99.9\% level ($\Delta \chi^2
\simeq 76$).
As evident from the residuals in Figure 2$d$, the two edges plus absorbed
power law model provides the best description of the soft X-ray spectrum.
Best-fit parameters are reported in Table 2 and confidence contours for
the edge energies versus optical depths are shown in Figure 3.
Assuming a conservative value of $\sim$ 10 eV (1$\sigma$) for the SIS
systematic uncertainties,
the edge at E $\simeq$ 0.70 $\pm 0.02$ keV is consistent with absorption
by OVI(671 eV), OVII(739 eV) and NVII(667 eV), where the energies quoted in
parenthesis are the edge energies in the rest frame (Lotz 1968).
The best agreement is with OVII. The second edge at E $\simeq 0.84 \pm0.02$
keV is consistent with OVIII(871 eV), only.
Inspection of the residuals in Figure 5 shows the presence of a
further edge-like feature at $\sim$ 1.25 keV, consistent with NeIX(1196 eV),
NeX(1362 eV) and Fe L edges of Fe ionization stages around FeXVIII.
Inclusion of a third absorption edge into the model is, however,
significant only at a $\sim$ 68 \% confidence level. Given its low statistical
significance, we do not consider it any further in the following.
It is interesting to note that both characteristic energies obtained for the
OVII and OVIII edges are redshifted by approximately the same amount
($\sim$ 25 eV, in the observer frame). We checked that this redshift is not
related to the Charge Transfer Inefficiency (CTI) of the SIS (see Otani \&
Dotani 1994 for details).

Because IC 4329A is a fairly bright point source, we may expect any systematic
problems due to remaining calibration uncertainties to be more
visible in IC 4329A than in other, weaker sources.
We point out that some features are present between $\sim$ 1.8 - 2.5 keV.
These are likely to be due to calibrations uncertainties in this
energy range (Si K-edge and Au M-edge of the detector). In this work, we will
assume that these features are not of cosmic origin, and are therefore
of no physical interest.

Also, even after applying our best-fit edge model (above),
there are still
residuals in the shape of an emission line around 0.6 keV and an
absorption feature around 0.5 keV.
In particular, the emission line can be fitted with a narrow
($\sigma \equiv 0$) gaussian line at E$\simeq$ 0.58 $\pm$ 0.05 keV
with EW $\simeq 30 \pm 25$ eV. The inclusion of this emission line
corresponds to a $\Delta \chi^{2}$ $\sim$ 4, which is significant
at more than 95\%.
Its energy is consistent with emission from OVII and/or OVIII. The
existence of similar emission lines is  predicted in warm absorber models
(Netzer 1993) and has been confirmed in the Seyfert 1 galaxy
NGC3783 (George, Turner \& Netzer 1995).  However,
it is also known from other ASCA observations that features around
0.5-0.6 keV are very sensitive to calibration uncertainties (due to the
O K-edge at $\sim$ 0.54 keV), background substraction and selection
criteria (Miura et al. 1995).
Therefore, we have carefully checked to establish the believability
of this soft emission line.

When different background regions are used and different selection criteria
(like BR\_EARTH $>$ 50 for ex.) are applied, both the absorption
and soft emission line features remain unchanged.
Using the ``gain'' command in XSPEC, we (linearly) shifted the energies of
SIS0 and SIS1 response matrices by $\pm$ 2 ADU ($\sim$ 7 eV), i.e.
within typical systematic errors of $\sim$ 10 eV (1$\sigma$)
for the DFE ``zero-level'' (Otani \& Dotani 1994).
After fitting the data again with the shifted matrices,
the absorption feature around 0.5 keV appears smoothed out. It is
therefore probably due to calibration uncertainties.  While the
soft emission line
changes in intensity it remains statistically significant in every case.
However, the emission line is significant only in the SIS1
data.  Also, it appears to go away when only the night data are
considered although, in this case, the statistics are considerably reduced.
We conclude that with the present data, the existence of this emission
line is doubtful.

\subsection{Overall spectrum}

We now include also the data from E $>$ 5 keV and proceed to the analysis
of the spectrum in the broad (0.4 - 10 keV) energy band.
One striking result is that we detect a clear hardening of the
spectrum even down to 5-6 keV, likely due to the presence of a reflection
component, as previously found by Mushotsky et al. (1995).
Figure 4 illustrates the existence of the hard tail and of a Fe K emission
line around 6.2 keV. To produce this Figure,
the spectrum was fitted between 0.4 - 5 keV with the two edges plus power law
model (section 3.1.2., Table 2) and the data above 5 keV were added back in.
The data were rebinned to more than 500 counts/bin in order to better show the
presence of the hard tail. The
systematic positive values of the residuals, before and after the FeK emission
line, are a clear indication that the continuum is increasing at higher
energies. This is also confirmed by the huge
($\sigma\ \sim$ 1.17 keV, EW $\sim$ 760 eV) values obtained for the Fe K line
(E $\sim$ 6.25 keV) if we include only a single gaussian line in the model,
which tries to account for the line plus the hard tail.
We underline that in this case, as well as in all the following spectral fits,
all the parameters of the two edges plus power law in the model are free.
Given the importance of the residuals above 7 keV, we carried a number of tests
to confirm the reality of the hard tail. The use of a background choosed
from the outer region of the relevant chip in the SIS had no significant effect
on either the overall spectrum or the strenght of the hard tail.
The use of different extraction radii for the source (from r=4 arcmin to
r=9 arcmin) and more conservative selection criteria (like BR$\_$EARTH$>$50 to
avoid light leakage) gave similar results as well.

In a recent ASCA long observation of the Seyfert 1 galaxy MCG-6-30-15, a
very broad iron line has been observed (Tanaka et al. 1995).
It has been well modelled by the sum
of a broad (E $\sim$ 5.6 keV, $\sigma \sim$ 0.54 keV, EW $\sim$ 150 eV) and
narrow (E $\sim$ 6.5 keV, $\sigma \sim$ 0.20 keV, EW $\sim$ 150 eV) gaussian
lines. These have been interpreted as the signature of the ``double-horned''
line expected from an accretion disk (e.g., Fabian et al. 1989). If a
double-gaussian parameterization is used for IC 4329A instead of a single line,
one gaussian converges to a very broad ($\sigma \sim$ 1.2 keV, EW $\sim$ 713
eV) line
centered at $\sim$ 6.2 keV and the other one to a narrow ($\sigma =$ 0,
EW $\sim$ 53 eV) at the same energy.
Since the data require an unusually broad gaussian that is non-physical and
inconsistent with any previous X-ray measurement, we
prefer to interpret this result as meaning that the high energy part of the
spectrum requires an additional continuum component.

The presence of a Fe K emission line and a hard tail
is usually interpreted in terms of
a model in which the primary continuum emission is reprocessed by optically
thick matter, i.e. a reflection component (Lightman \& White 1988,
Nandra \& Pounds 1994).
Therefore, we added a reflection component in the two edges plus
power law model. A gaussian emission line is also included in the fit.
The reflection used is the ``plrefl'' model in XSPEC and is calculated
following Lightman \& White (1988).
The reflecting matter is assumed to be thick, cold and with cosmic abundances
(Morrison and Mc Cammon 1983).
The inclination angle $i$ has been fixed to 0, i.e. the reflecting plane
is seen face-on, and the covering factor $\Omega/2 \pi$ has been fixed
to 1.
The source is assumed to emit isotropically and the spectrum is integrated over
all viewing angles.
The relative normalization R ( $\equiv$
A$_{\rm refl}$/A$_{\rm pl}$) between the reflected
component (A$_{refl}$) and the direct component (A$_{pl}$) is a free parameter.
This implies that the normalization of the reflection component is allowed
to be independent of the normalization of the power law.
The parameter R, alone, includes the uncertainties concerning the geometry
of the reflector (i.e., the covering fraction), the possibility that the
source is anisotropic (e.g., Ghisellini et al. 1991) and time-lag effects.
This implies, for example, that a best-fit value of R $\sim$ 1 can be explained
if the reprocessing material covers $\sim \ 2\pi$ of the source. If R $>$ 2,
then
the result cannot be explained {\bf only} in terms of a covering factor
increase,
and either the source must be anisotropic or there are time delay effects.

The photon index has been left free to vary.
The best-fitting parameters obtained with a reflection fit are
given in Table 3. The unfolded spectrum and residuals are shown in Figure 5.
The measured photon index of 1.98 $\pm$ 0.05, $N_{\rm H} \simeq$ (3.48 $\pm$
0.12)
$\times$ 10$^{21}$ cm$^{-2}$ and Fe K line parameters
(E $\simeq 6.20 \pm 0.07$ keV, $\sigma \simeq
0.11 \pm 0.08$ keV and EW $\simeq 89 \pm 33$ eV) are consistent with previous
$Ginga$ results (PYM90, Fiore et al. 1992).
Confidence contours in the Fe K parameter space $\sigma$ - E are
well constrained (Figure 6).
These show that the Fe K line is consistent with redshifted cold iron at more
than 90\% confidence level, and is marginally broad at an $\sim$ 80\%
confidence level.
As discussed later on, these characteristics of the iron line
suggest that we are dealing with a disk geometry (Fabian et al. 1989,
Mushotsky et al. 1995).

As already mentioned in section 3, the GIS spectral analysis is
completely consistent with that derived above for the SIS data.
In fact, the inclusion of a reflection component, assuming an inclination
$i$ = 0 and covering factor $\Omega/2\pi =$ 1 as for the SIS spectrum,
improves the GIS fit at a confidence level $>$ 99\% ($\Delta \chi^{2}\ \sim
21$).
Comparison of the GIS best-fit parameters (Table 1B) with those obtained
from the SIS data (Table 3), shows very good agreement from both detectors
and allow us to be fairly confident in the proposed best-fitting model.
The major discrepancies are in the $N_{\rm H}$ and photon index best-fit
values which are respectively higher and flatter in the GIS than in the SIS
spectrum.
However, these differences are likely attributed to the
fact that the GIS is not sensitive enough at soft energies to detect
the absorption edges. As they are not fitted explicitely in the GIS spectrum,
they likely enhance the absorption value and therefore contribute to
flatten the GIS spectrum.
This hypothesis is also supported by the fact that the fit of the
spectrum with all 4 detectors (SIS + GIS) gives results almost identical to
that found with the SIS only.


\subsection{On the model-dependence of photon index, reflection normalization
and Fe K line parameters}

With the above best statistical fit of IC 4329A
(Table 3), we found that there is a strong (linear) correlation between
the photon index and the amount of reflection (Figure 7). This relation
should be borne in mind when interpreting the results of this analysis.
A similar effect has been discussed for the analysis of
MCG-2-58-22 (Weaver et al. 1995).
While a high best-fit value of R may
indeed indicate that the true intrinsic photon index is steep, it also
depends strongly on fitting the correct amount of reflection, as well as
the shape of the reflected spectrum modelled. There is no guarantee that the
current reflection model is entirely correct and we may biasing the fits
with a slightly incorrect shape or normalization of the reflection.
Different analysis procedures may, therefore, lead to slightly different
results.
For example, fixing the photon index to its 90\% upper limit ($\Gamma
\simeq 1.92$) obtained from the fit in the 0.4 - 5 keV energy band (Table 2),
the amount of reflection converges to R $\simeq$ 2.1.
However, fixing the photon index from the fit at soft energies is not
necessarily correct in this case because we may also bias our fit with an
improper parameterization of the warm absorber.
Alternatively, fixing R $= 1$, a value appropriate for a face-on
slab subtending 2$\pi$ solid angle at the X-ray source, the fit gives
$\Gamma \simeq 1.89$. However, also in this case, there is no reason to fix R.
The best solution of the present data is therefore that obtained with $\Gamma$
and R
free to vary (Table 3).

Because of the correlation between photon index and the amount of reflection
in the model we investigated by how much this could affect the parameters
obtained for the Fe K line. For this, we calculated the 90\% confidence
contours of $\sigma$ versus energy for Fe K with R varying from zero
(no reflection) to 3.35 (the best-fit value).
These contours are shown in Figure 8 for R=0, 0.3, 1, 1.5, 2.1 and 3.35.
The corresponding best-fit values of the Fe K line width are 1.17, 0.89,
0.21, 0.15, 0.13 and 0.11 keV, respectively. The $\chi^2$ values are
641, 640, 639, 634, 630 and 626 for 547 d.o.f.
We see that the energy of the Fe K line depends only slightly on the
exact amount of reflection. These contours are indicative of 3 different
``solutions'' for the line width: (i) assuming little or no reflection
(R $\lsimeq$ 0.9) results in a very broad Fe K line with
characteristic width of $\gsimeq$ 0.8 keV and EW $\gsimeq$ 380 eV,
(ii) assuming an intermediate contribution from reflection (R $\sim$ 1 - 2)
results in two local minima, one near $\sigma \sim$ 0.6 - 0.8 keV
and the other near $\sigma \sim$ 0.1 - 0.2 keV, and
(iii) assuming a strong reflection component (R from $\sim$ 2 to the
best-fit value $\sim$ 3.3) results in one minimum at $\sigma \sim
0.11$ keV with well constrained contours.
Fitting simultaneously the 2 SIS and 2 GIS detectors yielded very similar
results. These contours, as well as the best-fit values, are similar to those
obtained by Mushotsky et al. (1995) although their
fit yields a slighlty larger Fe K width than inferred from our best-fit
(case (iii)).
The difference may be a consequence of the slightly different continuum
modelling, probably caused by the different energy band used by the authors.

To summarize, we find that although the photon index and reflection
normalization
are strongly coupled together, the most likely description of the spectrum
of IC 4329A is given by its best statistical fit (Table 3).
We have also shown that despite this intrinsic uncertainty, the Fe K parameters
depend only little on the exact amount of reflection, provided R is
higher than $\sim$ 2, as required by the best-fit model.
With these caveats borne in mind, we can discuss and try to interpret
these results physically.

\section{Discussion}

Above results show that the best interpretation of the soft X-ray spectrum
of IC 4329A is in terms of photoionized absorber.
Therefore, we have also fitted the SIS data between 0.4 and 5 keV
with a warm absorber model, also including a uniform cold absorber.
The model has been produced using the photoionization code CLOUDY (Ferland
1991)
and using the assumptions as in Fabian et al. (1994).
Free parameters are the ionization parameter $\xi = {\rm L}/nR{^2}$
(erg s$^{-1}$ cm), the ``warm'' column density $N_{\rm W}$ (cm$^{-2}$),
and the cold column density $N_{\rm H}$ (cm$^{-2}$).
The photon index has been fixed to 2.0, a value close to the SIS best-fit
value.
The fit yields an acceptable fit with $\chi^{2}_{red}/$d.o.f. = 1.23/480.
Although it is statistically better than with the blackbody model, it is
worse than with the two absorption edges.
We attribute this difference to the fact that the model used is based
on simple basic assumptions which may not reflect the
reality, namely: a thin geometry for the photoionized gas, solar
abundances, constant temperature, constant density and a source
luminosity of $10^{43}$ erg s$^{-1}$.
Further, the model does not include the emission lines expected
from a more realistic warm absorber (Netzer 1993).
We obtain log $\xi \simeq$ 1.12 $\pm$ 0.04, log $N_{\rm W} \simeq$ 21.70 $\pm$
0.02 and log $N_{\rm H} \simeq$ 21.49 $\pm$ 0.01.
The fitted neutral column density is again about one order
of magnitude higher than the Galactic value.

Interesting is to note that with the 2 edges parameterization of the warm
absorber, both edges best-fit energies are redshifted by approximately
25 eV. If real, this redshift correspond to a Doppler velocity of about 10000
km
s$^{-1}$. Although this result depends strongly
on the correct modelling of the warm absorber (e.g., the redshift of the lower
edge may be interpreted as a blend of OVI and OVII edges) and on the
precise calibration of the energy scale, it may be the signature of infalling
matter or largescale turbulance associated to the warm absorber region.
Firm conclusions on this point require more investigations, but we point
out that edge energies shifted compare to their expected energy have been
reported also for the observations of MCG-6-30-15 and NGC4051 (Otani 1995).

As suggested by Krolik \& Kriss (1995), resonance line scattering
may significantly contribute to the total opacity of the warm gas. It is worth
noting therefore that it is a potential confusing factor when interpreting
the energy and optical depth of the absorption edges.
For example, the presence of blended resonance oxygen lines may cause a
misidentification of OVIII as OVII due to the apparent energy
shift in the onset of the OVIII edge. Alternatively, it could contribute to
shift the apparent energies of the absorption edges.
Unfortunately, quantitative considerations on resonance absorption being
strongly model-dependent (e.g. on the covering factor and velocity of the warm
gas), and given the limited spectral resolution of the SIS, we cannot
exclude the presence of resonance absorption lines in the present data.

A striking alternative to the warm (plus cold) absorber
is absorption by cold matter with abundances free to vary.
IC 4329A is known to exhibit absorption of a few 10$^{21}$ cm$^{-2}$.
This absorption is consistent with the fact that its host galaxy
is seen edge-on, and so we examine whether the absorption edges
seen in the ASCA data can be due to neutral interstellar matter.
Since the K-edge energies of OVII, OVIII and NeX are close to
the L-edge energy of FeI, K-edge energies of NeI and MgI, respectively,
it is possible that abundances larger than
the cosmic value can mimic the absorption
edges normally interpreted as the presence of a warm absorber.
In fact, a fit of the soft energy band (0.4 - 5 keV) with a cold
absorber with variable abundances gives
an equally good fit as that obtained with our previous
parameterization using two warm edges (section 3.1.2).
However, extremely high abundances ($\sim$ 5 - 15 times the cosmic value)
of all metals are required.
Also, known calibration problems around 0.5 - 0.6 keV (see section 3.1.2)
make the results from this fit doubtful.
A deeper investigation using these and other data is currently
being conducted and will be presented in a
future paper (Hayashida et al., in preparation).

The X-ray spectrum of IC 4329A also requires a remarkably strong
reflection component.
In a recent re-analysis of $HEAO-1$ (2 - 40 keV) data, a similar
amount (R $\sim$ 2.3 $\div$ 4.2) of reflection component has been
detected in IC 4329A (Weaver, Arnaud, \& Mushotzky 1995).
Similarly, we find in the ASCA spectra $\sim$ 2 - 3 times more reflection
relative to the continuum than in the $Ginga$ observation (PYM90,
Fiore et al. 1992, Nandra \& Pounds 1994). The present data also suggest
a slight, but significant, increase in the absolute amount of reflection.
Although we might recall that the ASCA best-fit spectra,
as well as the $HEAO - 1$ and $Ginga$ spectra, exhibit a strong coupling
between photon index and amount of reflection (see section 3.3), these
results indicate the presence of a strong and variable reflection component.
It should be noted that a time-resolved analysis of the $Ginga$
observation also suggested day to day variations in the amount of reflection
component in IC 4329A (Fiore et al. 1992).

If the geometry of AGN consists on a source
illuminating hard X-rays above and below a soft X-ray emitting disk
(e.g. Haardt, Maraschi \& Ghisellini 1994), and these hard X-rays are
produced by Inverse Compton scattering of the soft X-rays, then
the radiation field within the hard X-ray emitting
region is likely to be strongly anisotropic (Ghisellini et al. 1991).
The reflecting matter may therefore see more radiation than that emitted
towards the observer, by up to a factor of $\sim$ 5, as a direct
consequence of the electrons losing more energy on head-on collisions with
the soft X-ray photons than on collisions from behind.
In that case, a strong reflection component becomes physically plausible.
However, anisotropy alone is unlikely to explain the increased
contribution of the reflected component compared to that of the direct
continuum because it requires a change in the anisotropic
factor. In other words, anisotropy can explain a high value of R
but not a variation of R.  Instead,
a time lag effect could, alone, explain all of the observational constraints.
Though IC 4329A has never exhibited strong, rapid variability, it has varied up
to
a factor of $\sim$ 2.8 (from a 2-10 keV flux of $\sim 5 \times 10^{-11}$ erg
cm$^{-2}$ s$^{-1}$ to $\sim 14 \times 10^{-11}$ erg cm$^{-2}$ s$^{-1}$) in a
timescale of $\sim$ 5-6 years, with a doubling timescale down to $\sim$ 1 year,
in its $\sim$ 20 years history of X-ray observations.
A time lag could therefore cause strong amount of reflection.
Our results can indeed be understood in this context if the continuum flux
decreased by some amount prior to the ASCA observation.
Specifically, the flux would have to have decreased by a factor of
$\sim$ 2 from a time when it was
higher than that measured by $Ginga$, and the reflection
component could not have had time to respond to this decrease.
This can simultaneously explain a high value of R, a variation of R, and a
higher absolute amount of reflection as compared to $Ginga$.


However, another difficulty with the above results is that strong
reflection should be accompanied, in most cases, by a strong iron line.
We note that the reflected spectrum used in our fitting procedure
(section 3.2) is averaged over all viewing angles and has accordingly
a slightly different shape than a face-on reflected spectrum.
The best-fit amount of reflection R $\sim$ 3.3 corresponds therefore to a
($\sim$ 1.33 times) lower value R $\sim$
2.5 for a face-on view (Ghisellini, Haardt \& Matt 1994, Magdziarz \&
Zdziarski 1995). As a rough estimate, assuming an Fe K line EW of $\sim$ 140 eV
in the case R=1 for a cold face-on disk and an incident power law spectrum
with $\Gamma \sim 2.0$, we expect an EW of $\sim$ 300 eV. The best-fit
amount of reflection requires therefore $\sim$ 3 times more line than
observed (EW $\simeq$ 89 $\pm$ 33 eV). This difference may be reconciled
by reducing the abundance of iron, or by increasing the abundance of the light
metals (lighter than iron) (Reynolds, Fabian \& Inoue 1995). Alternatively,
if the Fe K line arises in an ionized disk, its EW can be very low even in the
presence of strong reflection because of either resonant trapping of line
photons or complete ionization of iron (Ross \& Fabian 1993, \.Zycki \& Czerny
1994).
Compare to the neutral case, the ionized disk has lower opacity at lower
energies. This might help to explain why the reflection shows up at such low
($\sim$ 5-6 keV) energies in this source.
However, we show below that the above constraints can be considerably reduced
because the EW obtained with an accretion disk line model is significantly
higher.

The parameters of the Fe K line obtained from the SIS spectral
analysis are consistent with the line being redshifted by $\sim$ 100 eV and
slightly broad ($\sigma \sim 110$ eV).
These support the idea that a fit of the line with a disk line model
(Fabian et al. 1989) would be more appropriate than a gaussian fit.
A disk line model takes into account the effects of doppler-broadening and
gravitational and transverse redshifts in order to calculate the
characteristic line profile. The resulting line profile is highly
asymmetric, and often ``double-horned''. A remarkable example of the
applicability of this model to Seyfert 1 galaxies has come from
ASCA long observations of MCG-6-30-15 (Tanaka et al. 1995).

We attempted a fit of the SIS data with the ``diskline'' model
available in the XSPEC program. This model calculates the intensity and
profile of the line from a flat, optically thick accretion disk as
calculated in Fabian et al. (1989). The details of the line
depend on the inclination $i$ of the disk, the inner R$_{in}$ and
outer R$_{out}$ disk radii, and on the line emissivity $\alpha$ (assumed to
vary as R$^{-\alpha}$).
Following Mushotsky et al. (1995), we fix R$_{in}$ = 3 R$_{\rm s}$ and
R$_{out}$ = 1000 R$_{\rm s}$, where R$_{\rm s}$ is the Schwarzschild
radius of the black hole (R$_{\rm s} \sim 3\times 10^{5} {\rm M/M}_{\odot}$ cm,
with
M the mass of the black hole).
The inclination of the disk, $i$, is also fixed at 0 (face-on)
to be consistent with our assumption of a face-on reflection component.
The line energy is fixed
to 6.3 which correspond to the redshifted energy of a fluorescence Fe K
line. Only the normalization of the line and its emissivity $\alpha$ are
free to vary. The fit yields
$\alpha \simeq$ 2.30 $\pm$ 0.20 and an EW of 188 $\pm$ 60 eV.
If $i$ is allowed to vary, we obtain an upper limit of $i\ \lsimeq$ 25
degrees which confirms a posteriori that the assumption of a disk face-on is
substantially correct. In this case, the EW is enhanced to
$213_{-63}^{+76}$ eV.
These results are characteristic of a disk observed face-on, in which
most of the fluorescence Fe K line is emitted within a few tens of
Schwarzschild radii.
As stated above, with the use of a disk line model, the Fe K EW
is in rough agreement with the value expected from the reflection
component.

Strictly speaking, if a disk line model is included, we should also take into
account relativistic effects on the (continuum) reflection component in order
to have a self-consistent model. This model is not available at the moment.
This problem will be explored in a later paper.
However, relativistic corrections should produce merely
an energy shift and/or broadening of the reflection component.
Therefore, we expect these shouldn't significantly affect the above results.

It should be noted however that the velocity width (FWHM $\simeq 12000 \pm
8500$
km s$^{-1}$) of the Fe K line is still consistent with an average derived
FWHM of $\simeq 4070$ $\pm$ 30 km s$^{-1}$ for the strong Balmer lines
(H$_{\alpha}$, H$_{\beta}$ and H$_{\gamma}$) which dominate the optical
spectrum of IC 4329A.
This suggests that the Fe K line might, alternatively, be produced from the
broad line region. One-sided beaming of the continuum emission and/or
obscuration may be necessary, however, to explain the apparent redshift
of the line.

\section{Conclusions}

To summarise, the present observation has revealed the following.

(i) The apparent soft X-ray spectrum of IC 4329A is complex with a predominant
soft-excess (compared to a simple absorbed power law) at E $\lsimeq$ 0.8 keV
and deep absorption structures between $\sim$ 0.8 - 1.2 keV.
The unprecedent high ($>$ 10 in every channel) signal to noise ratio
of this observation allows us to discard, directly from the residuals,
a blackbody emission model as an alternative explanation for the soft excess.
The soft X-ray spectrum is instead best described by two
absorption edges at E $\sim$ 0.70 keV and E $\sim$ 0.84 keV,
consistent with absorption by OVI, OVII or NVII, and OVIII, respectively.
This complex absorption is clearly the signature of a warm absorber,
confirming and extending previous results obtained from ROSAT
observations (Madejski et al. 1995).

(ii) Confirmation of the existence of a ``hard tail'' at higher energies,
characteristic of a reflection component (Mushotsky et al. 1995). The hard
tail is clearly detected in both GIS and SIS, at similar intensities.
When the reflection is included, we find that the data prefer a stronger
absolute
amount of reflection and stronger contribution relative to the direct continuum
than previously observed. A time lag effect can account for all observational
constraints.

(iii) Because of the presence of the reflection component, the Fe K line width
and
equivalent width are model-dependent. These are, however, only weakly
model-dependent provided R is higher than $\sim$ 2.1 (section 3.3).
With the proposed best-fit model (with R $\simeq$ 3.3 $\pm$ 1.1),
we find evidence that the Fe K line is redshifted (by $\sim$ 100 eV) and
slightly broad ($\sigma \simeq 0.11 \pm 0.08$ keV), as expected if the emission
line
is emitted by an accretion disk. The weak (EW $\simeq$ 89 $\pm$ 33 eV) Fe K
(gaussian) line requires iron underabundance or/and ionized matter to be
consistent with strong reflection. A fit with a diskline model which includes
the wings of the line can, however, account for most of the discrepancy as
well.

\section{Acknowledgements}

\ms M.C. thanks K. Leighly and S. Molendi for helpful discussions.
M.C. also acknowledges financial support from the Science and
Technology Agency of Japan (STA fellowship), hospitality from the
RIKEN Institute and support from the European Union.
The authors also thank an anonymous referee for valuable comments.

\vfill\eject

\section{References}

\pn
Arnaud, K.A., Haberl, F., \& Tennant, A., 1991, XSPEC User's Guide, ESA TM-09

\pn
Elvis, M., Lockman, F.J., \& Wilkes, B.J., 1989, AJ, 97, 777

\pn
Fabian, A.C., Rees, M.J., Stella, L. \& White, N.E., 1989, MNRAS, 238, 729

\pn
Fabian, A.C. et al., 1994, PASJ, 46, L59

\pn
Ferland, G.L., 1991, Ohio State Univ., Astronomy Dept. Internal Report 91-01

\pn
Fiore, F., Perola, G.C., Matsuoka, M., Yamauchi, M., \& Piro, L., 1992, A\&A,
262, 37

\pn
George, I.M., Turner, T.J., \& Netzer, H., 1995, ApJL, 438, L67

\pn
George, I.M., \& Fabian, A.C., 1991, MNRAS, 249, 352

\pn
Ghisellini, G., George, I.M., Fabian, A.C., \& Done, C., 1991, MNRAS, 248, 14

\pn
Ghisellini, G., Haardt, F., \& Matt, G., 1994, MNRAS, 267, 743

\pn
Guainazzi, M., Matsuoka, M., Piro, L., Mihara, T., \& Yamauchi, M., 1994, ApJ,
436, L35

\pn
Guainazzi, M., et al. 1995, in preparation

\pn
Guilbert, P.W., \& Rees, M.J., 1988, MNRAS, 233, 475

\pn
Haardt, F., Maraschi, L., \& Ghisellini, G., 1994, ApJL, 432, L95

\pn
Hayashida, K., et al., 1995, in preparation

\pn
Krolik, J.H., \& Kriss, G.A., 1995, ApJ, 447, 512

\pn
Leighly, K., et al., 1995, submitted to ApJ

\pn
Lightman, A.P. \& White, T.R., 1988, ApJ, 335, 57

\pn
Lotz, W., 1968, J. Opt. Soc. Am., 58, 915

\pn
Madejski, G.M., et al. 1995, ApJ, 438, 672


\pn
Magdziarz, P., \& Zdziarski, A.A., 1995, MNRAS, in press

\pn
Matsuoka, M. 1994, in ```New Horizon of X-ray Astronomy'', ed. F. Makino
\& T. Ohashi, (Tokyo: Universal Academy Press), p305

\pn
Mihara, T., Matsuoka, M., Mushotsky, R.F., Kunieda, H., Otani, C., Miyamoto,
S.,
\& Yamauchi, M., 1994, PASJ, 46, L137

\pn
Miura, N., et al. 1995, in preparation

\pn
Miyoshi, S., et al., 1988, PASJ, 40, 127

\pn
Morrison, R., \& McCammon, D., 1983, ApJ, 270, 119

\pn
Mushotsky, R.F., Fabian, A.C., Iwasawa, K., Kunieda, H., Matsuoka, M., Nandra,
K., \&
Tanaka, Y., 1995, MNRAS, 272, L9

\pn
Nandra, K., \& Pounds, K.A., 1994, MNRAS, 268, 405

\pn
Netzer, H., 1993, ApJ, 411, 594

\pn
Otani, C., 1995, PhD Thesis

\pn
Otani, C., \& Dotani, T., 1994, Asca News n.2, 25

\pn
Piro, L., Yamauchi, M., \& Matsuoka, M., 1990, ApJ, 360, L35 (PYM90)

\pn
Petre, R., Mushotsky, R.F., Krolik, J., \& Holt, S.S., 1984, ApJ, 280, 499

\pn
Ptak, A., Yaqoob, T., Serlemitsos, P.J., Mushotsky, R.F., \& Otani, C., 1994,
ApJ, 436, L31

\pn
Raymond, J.C. \& Smith, B.W., 1977, ApJS, 35, 419

\pn
Reynolds, C.S., Fabian, A.C. \& Inoue, H., 1995, MNRAS, in press

\pn
Ross, R.R., \& Fabian, A.C., 1993, MNRAS, 261, 74

\pn
Tanaka, Y., Inoue, H., \& Holt, S.S., 1994, PASJ, 46, 37

\pn
Tanaka, Y. et al., 1995, Nature, 375, 659

\pn
Turner, T.J., Weaver, K.A., Mushotsky, R.F., Holt, S.S., \& Madejski, G.M.,
1991, ApJ, 381, 85

\pn
Turner, T.J., \& Pounds, K.A., 1989, MNRAS, 240, 833

\pn
Weaver, K.A., Arnaud, K.A., \& Mushotzky, R.F. 1995, ApJ, in press

\pn
Weaver, K.A., Nousek, J., Yaqoob, T., Hayashida, K., \& Murakami, S., 1995,
ApJ, submitted

\pn
Wilkes, B.J., \& Elvis, M., 1987, ApJ, 323, 243

\pn
Wilson, A.S., \& Penston, M.V., 1979, ApJ, 232, 389

\pn
\.Zycki, P.T., \& Czerny, B., 1994, MNRAS, 266, 653

\vfill\eject

\bs
\centerline{TABLE 1 - GIS2+3 Data (0.7 - 10 keV)}
\centerline{A. Results of Power-Law + Fe K Line Fits}
\begin{center}
\begin{tabular}{ccccccc}
\hline
\hline
\multicolumn{1}{c}{$N_{\rm H}$}&
\multicolumn{1}{c}{$\Gamma$} &
\multicolumn{1}{c}{A$_{\rm pl}^a$} &
\multicolumn{1}{c}{E (FeK)} &
\multicolumn{1}{c}{$\sigma$ (FeK)} &
\multicolumn{1}{c}{EW (FeK)} &
\multicolumn{1}{c}{$\chi^{2}_{red}/d.o.f.$} \\
\multicolumn{1}{c}{($10^{21}$ cm$^{-2}$)} &
\multicolumn{1}{c}{} &
\multicolumn{1}{c}{} &
\multicolumn{1}{c}{(keV)} &
\multicolumn{1}{c}{(keV)} &
\multicolumn{1}{c}{(eV)} &
\multicolumn{1}{c}{} \\
\hline
$ 3.78_{-0.12}^{+0.18} $&$ 1.79_{-0.03}^{+0.02} $& $27.6_{-0.7}^{+0.7}$&$
6.19_{-0.15}^{+0.12} $&$ 0.53_{-0.24}^{+0.24} $&$ 233_{-73}^{+95} $ & 1.33/564
\\
\hline
\hline
\end{tabular}
\end{center}

\bs

\centerline{ }
\centerline{B. Results of Power Law + Fe K Line + Reflection Fits}
\vspace{-0.8truecm}
\begin{center}
\begin{tabular}{cccccccc}
\hline
\hline
\multicolumn{1}{c}{$N_{\rm H}$} &
\multicolumn{1}{c}{$\Gamma$} &
\multicolumn{1}{c}{A$_{\rm refl}^{a}$} &
\multicolumn{1}{c}{R$^b$} &
\multicolumn{1}{c}{E (FeK)} &
\multicolumn{1}{c}{$\sigma$(FeK)} &
\multicolumn{1}{c}{EW (FeK)} &
\multicolumn{1}{c}{$\chi^{2}_{red}/d.o.f.$} \\
\multicolumn{1}{c}{(10$^{21}$ cm$^{-2})$}&
\multicolumn{1}{c}{} &
\multicolumn{1}{c}{} &
\multicolumn{1}{c}{} &
\multicolumn{1}{c}{(keV)} &
\multicolumn{1}{c}{(keV)} &
\multicolumn{1}{c}{(eV)} &
\multicolumn{1}{c}{} \\
\hline
$4.22_{-0.27}^{+0.24}$ &$ 1.89_{-0.04}^{+0.05}$  & 73$^{+28}_{-28}$ &
$2.43_{-0.95}^{+1.04}$ &$ 6.22_{-0.08}^{+0.08} $ & $\lsimeq$ 0.26&
$78_{-24}^{+38} $& 1.30/563 \\
\hline
\hline
\end{tabular}
\end{center}
\pn
\pn
Note: Intervals are at 90\% confidence for one interesting parameter.

$ ^{a}$ Normalization in 10$^{-3}$ photons cm$^{-2}$ s$^{-1}$ keV$^{-1}$ at 1
keV.

$ ^{b}$ R $\equiv$ A$_{\rm refl.}$/A$_{\rm pl}$ (see section 3.2 for details)

\bs
\bs
\bs
\centerline{TABLE 2 - SIS0+1 Data (0.4 - 5 keV)}
\centerline{Results of Two Edges + Power Law Fits}
\begin{center}
\begin{tabular}{cccccccc}
\hline
\hline
\multicolumn{1}{c}{E$_{1}$} &
\multicolumn{1}{c}{$\tau_{1}$} &
\multicolumn{1}{c}{E$_{2}$} &
\multicolumn{1}{c}{$\tau_{2}$} &
\multicolumn{1}{c}{$N_{\rm H}$} &
\multicolumn{1}{c}{$\Gamma$} &
\multicolumn{1}{c}{A$_{\rm pl}^a$} &
\multicolumn{1}{c}{$\chi^{2}_{red}/d.o.f.$} \\
\multicolumn{1}{c}{(keV)} &
\multicolumn{1}{c}{} &
\multicolumn{1}{c}{(keV)} &
\multicolumn{1}{c}{} &
\multicolumn{1}{c}{(10$^{21}$ cm$^{-2}$)} &
\multicolumn{1}{c}{} &
\multicolumn{1}{c}{} &
\multicolumn{1}{c}{} \\
\hline
$ 0.70_{-0.03}^{+0.01} $&$ 0.63_{-0.08}^{+0.09} $&$ 0.84_{-0.02}^{+0.01} $
&$ 0.39_{-0.07}^{+0.06} $ & $ 3.30_{-0.09}^{+0.08} $&$ 1.89_{-0.03}^{+0.03} $&
29.5$_{-0.7}^{+0.7}$&1.15/477  \\
\hline
\hline
\end{tabular}
\end{center}
\pn
Note: Intervals are at 90\% confidence for one interesting parameter.

$ ^{a}$ Normalization in 10$^{-3}$ photons cm$^{-2}$ s$^{-1}$ keV$^{-1}$ at 1
keV.

\vfill\eject

\bs
\centerline{TABLE 3 - SIS0+1 Data (0.4 - 10 keV)}
\centerline{Results of Two Edges + Power Law + Fe K Line + Reflection Fits}
\vspace{-1.4truecm}
\begin{center}
\begin{tabular}{cccccccccccc}
\hline
\hline
\multicolumn{1}{c}{E$_{1}$} &
\multicolumn{1}{c}{$\tau_{1}$} &
\multicolumn{1}{c}{E$_{2}$} &
\multicolumn{1}{c}{$\tau_{2}$} &
\multicolumn{1}{c}{$N_{\rm H}$} &
\multicolumn{1}{c}{$\Gamma$} &
\multicolumn{1}{c}{A$_{\rm refl}^{a}$} &
\multicolumn{1}{c}{R$^b$} &
\multicolumn{1}{c}{E (FeK)} &
\multicolumn{1}{c}{$\sigma$(FeK)} &
\multicolumn{1}{c}{EW (FeK)} &
\multicolumn{1}{c}{$\chi^{2}_{red}/d.o.f.$} \\
\multicolumn{1}{c}{(keV)} &
\multicolumn{1}{c}{} &
\multicolumn{1}{c}{(keV)} &
\multicolumn{1}{c}{} &
\multicolumn{1}{c}{(10$^{21}$ cm$^{-2})$}&
\multicolumn{1}{c}{} &
\multicolumn{1}{c}{} &
\multicolumn{1}{c}{} &
\multicolumn{1}{c}{(keV)} &
\multicolumn{1}{c}{(keV)} &
\multicolumn{1}{c}{(eV)} &
\multicolumn{1}{c}{} \\
\hline
$ 0.70_{-0.02}^{+0.01} $&$ 0.67_{-0.09}^{+0.07} $&$ 0.84_{-0.02}^{+0.01} $&$
0.40_{-0.07}^{+0.07}$& $3.48^{+0.12}_{-0.13}$ &$ 1.98_{-0.05}^{+0.05} $ &
$102^{+41}_{-34}$ &
3.35$_{-1.10}^{+1.17}$&$ 6.20_{-0.06}^{+0.07} $ &$ 0.11_{-0.09}^{+0.07} $&$
89_{-32}^{+34} $& 1.15/546 \\
\hline
\hline
\end{tabular}
\end{center}
Note: Intervals are at 90\% confidence for one interesting
parameter.

$ ^{a}$ Normalization in 10$^{-3}$ photons cm$^{-2}$ s$^{-1}$ keV$^{-1}$ at 1
keV.

$ ^{b}$ R $\equiv$ A$_{\rm refl.}$/A$_{\rm pl}$ (see section 3.2 for details)

\vfill\eject

\centerline{\bf Figure captions}
\par\noindent
{\bf Figure 1}: Residuals of the SIS spectrum between 0.4 - 10 keV, for a model
consisting of a single absorbed power law with $\Gamma \sim 1.75$. This
figure illustrates the soft excess, absorption features around 1 keV, Fe K
line at $\sim$ 6.2 keV and hard tail starting at E $\gsimeq$ 5 keV (see also
Fig. 4).
\ms
{\bf Figure 2}: Residuals of the SIS spectrum between 0.4 - 5 keV for a model
consisting of: ($a$) absorbed blackbody plus power law, ($b$) absorbed
power law plus Raymond Smith spectrum covered by only the Galactic column
density,
($c$) absorption edge plus absorbed power law and ($d$) two absorption edges
plus absorbed power law.
\ms
{\bf Figure 3}: Contour plots illustrating the (observed) energy and optical
depth
68, 90 and 99 per cent confidence levels of the two absorption
edges at ($a$) E $\sim$ 0.7 keV and ($b$) E $\sim$ 0.84 keV. The
redshifted energies of the absorption edges have been marked in the figures.
\ms
{\bf Figure 4}: Residuals of the SIS spectrum between 0.4 - 10 keV, for a model
consisting of a two absorption edges plus an absorbed power law fitted
between 0.4 - 5 keV. The data have been rebinned with more than 500 counts/bin
to better show the systematic increase of the residuals above 5 keV.
\ms
{\bf Figure 5}: SIS0+1 unfolded spectrum, best-fitting power law plus
two absorption edges plus Fe K line plus reflection model (upper panel).
Residuals in the form of the ratio (data/model) are shown in the lower panel.
\ms
{\bf Figure 6}: Fe K width and energy 68, 90 and 99 per cent confidence contour
levels for the best fit model, as given in Table 3.
\ms
{\bf Figure 7}: Contours at 68, 90 and 99 per cent confidence of amount of
reflection (R) vs. photon index ($\Gamma$). They illustrate the strong
correlation between the two parameters.
\ms
{\bf Figure 8}: Contours at 90 per cent confidence of amount of line width,
$\sigma$, vs. observed line energy calculated for various amounts of
reflection.
The best-fit values are marked with crosses.
\end{document}